\documentclass[letterpaper]{article}
\usepackage{spconf,amsmath,graphicx}            
\usepackage{amssymb,booktabs,enumitem}          
\usepackage[hidelinks]{hyperref}

\title{Revisiting Input Time-frequency Representations in\\ Multi-pitch Estimation for Vocal Ensembles}

\name{Junyoung Koh$^{1,2}$ \qquad Hao-Wen Dong$^{2}$}
\address{$^{1}$Yonsei University \quad $^{2}$University of Michigan \\
  \small \texttt{junyounk@umich.edu} \quad \texttt{hwdong@umich.edu}}

\begin{document}
\ninept
\renewcommand{\footnotesize}{\normalsize}
\maketitle

\begin{abstract}
Multi-pitch estimation in vocal ensembles is challenging because singers occupy overlapping pitch ranges and often sing at closely spaced fundamental frequencies, causing their harmonics to overlap in time-frequency representations.
Existing models commonly use harmonic constant-Q transform (HCQT)-based representations to provide frequency-adaptive resolution, at the cost of expensive feature extraction when training mixtures are generated on the fly.
We revisit this design and compare HCQT with a linear short-time Fourier transform (STFT), whose frequency bins are directly provided as model inputs.
Despite its fixed frequency resolution and the absence of a pitch-aligned input grid, the linear STFT outperforms HCQT while substantially reducing feature-extraction cost.
Further analysis shows that a longer analysis window or broader spectral coverage provides no additional improvement, while restricting the input to the predicted pitch range reduces the advantage of the linear STFT.
These results suggest that finer frequency resolution does not necessarily improve vocal-ensemble MPE, and that shorter analysis windows can be more effective for time-varying vocal pitches\footnote{
Code and configurations are at \url{https://github.com/Junst/stft-vocal-mpe}}.
\end{abstract}


\section{Introduction}
\label{sec:intro}

Multi-pitch estimation (MPE) identifies every fundamental frequency active in each frame of a polyphonic signal.
In vocal ensembles multi-pitch estimation, closely spaced fundamental frequencies can produce overlapping harmonic components, making their separation in time-frequency representations difficult.

Current vocal-ensemble MPE models commonly use the harmonic constant-Q transform (HCQT)~\cite{bittner2017deep}.
The constant-Q transform (CQT)~\cite{brown1991cqt} uses frequency-dependent analysis windows to maintain approximately constant relative frequency resolution.
Stacking CQTs computed at harmonic ratios further aligns the harmonics of each fundamental across channels, allowing convolutional filters to aggregate harmonic information across pitches.
These properties, however, increase feature-extraction cost, particularly when training mixtures are generated on the fly from isolated stems.

\begin{figure}[t]
\centering
\includegraphics[width=\columnwidth]{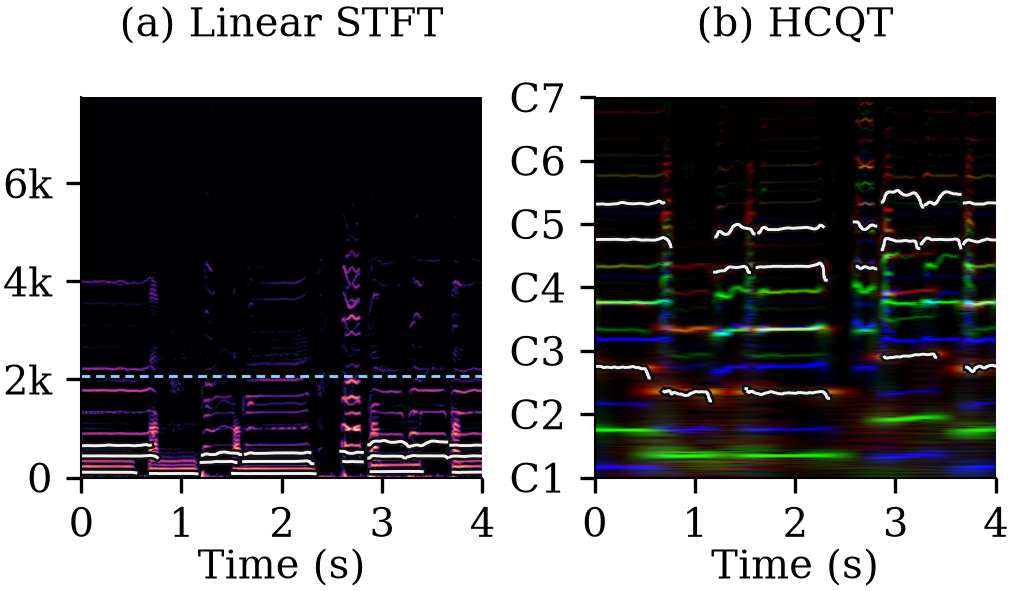}
\caption{
\textbf{The two representations compared, on a four-second a cappella recording in jaCappella~\cite{yoshii2023jacappella}: (a) the linear STFT with 2048-sample window and (b) the harmonic CQT.}
Reference fundamental frequencies are shown in white, with $h=1,2,3$ shown in red, green, and blue, respectively, in (b). The dashed line in (a) marks C7, the upper bound of the predicted pitch range.
}
\label{fig:concept}
\end{figure}

In this paper, we revisit the HCQT design and study the \textit{linear short-time Fourier transform (STFT)} as an input representation for vocal-ensemble MPE. A linear STFT provides the first 360 bins of a single fixed-window STFT directly to the model, without mapping them onto log-pitch coordinates. Unlike the HCQT, this representation has fixed frequency resolution, does not use frequency-dependent analysis windows, and has no pitch-aligned input grid.

To evaluate the effectiveness of the linear STFT for vocal-ensemble MPE, we conduct experiments on a range of choir and a cappella datasets, including jaCappella~\cite{yoshii2023jacappella}, ACappellaSet~\cite{pan2025acappellaset}, Korean Multi-Singer~\cite{aihub2021multisinger}, Cantor\'{\i}a~\cite{cuesta2022cantoria}, and ChoralSynth~\cite{narang2023choralsynth}.
We compare the linear STFT and HCQT representations across five encoder architectures.
The linear STFT consistently outperforms HCQT-based inputs while substantially reducing feature-extraction time.
We further show that increasing the analysis-window length or spectral range does not improve overall performance, whereas limiting the input to the predicted pitch range reduces the advantage of the linear STFT.

\begin{figure*}[t]
\centering
\includegraphics[width=\textwidth]{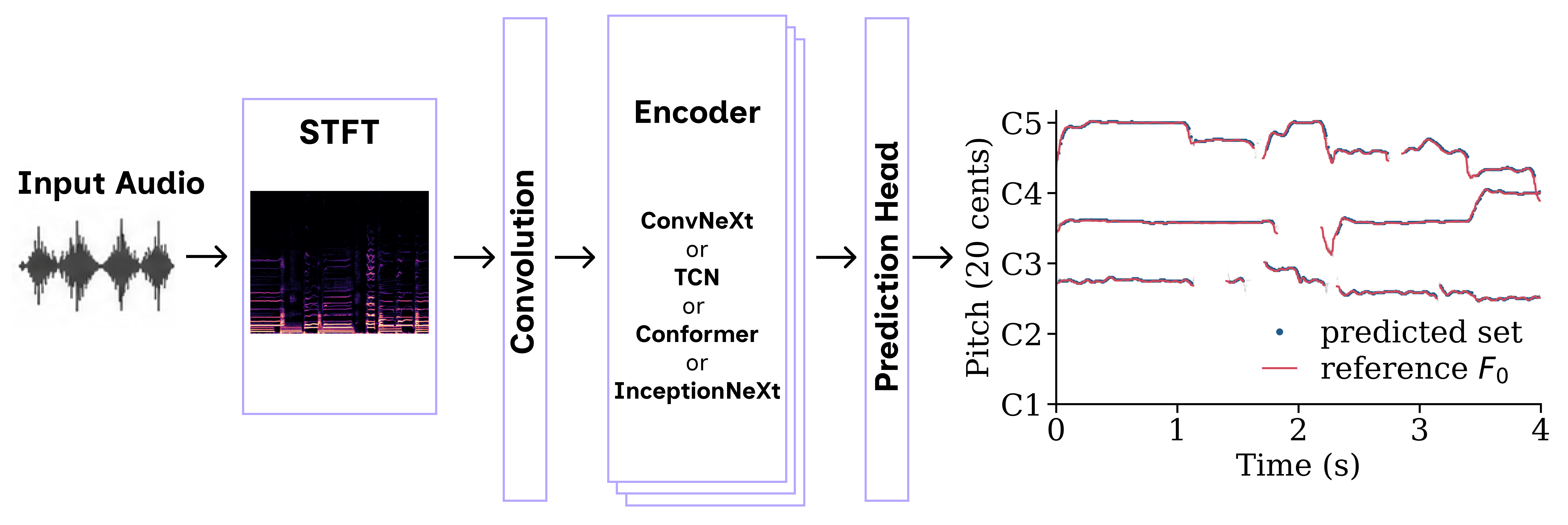}
\caption{
\textbf{Model architecture.}
The first 360 bins of a 2048-sample STFT (0--7.7\,kHz) are processed by a convolutional stem and an encoder, followed by a frame-wise prediction head over 360 pitch bins (C1--C7).
}
\label{fig:arch}
\end{figure*}

\section{Related Work}
\label{sec:related}

\vspace{0.15em}
\noindent\textbf{Vocal-ensemble multi-pitch estimation.}
Classical multi-pitch estimation relied on harmonicity and spectral smoothness~\cite{klapuri2003multiple} or melody-oriented salience~\cite{salamon2014melody}, while early a cappella transcription used latent component analysis over singer templates~\cite{schramm2017msingers}.
Neural transcription models include Onsets and Frames~\cite{hawthorne2018onsets}, Basic Pitch~\cite{bittner2022lightweight}, Cerberus~\cite{manilow2020simultaneous}, and deep CNN-based MPE~\cite{weiss2022comparing}.
Vocal-ensemble methods instead commonly follow the salience-map formulation of Deep Salience~\cite{bittner2017deep}, predicting a time-by-pitch activation matrix.
The Late/Deep CNN~\cite{cuesta2020multif0} applies this formulation to vocal quartets using an HCQT input, while subsequent work assigns estimated pitches to SATB parts~\cite{cuesta2022voiceassign} or separates the ensemble before transcription~\cite{petermann2020choir}.

\vspace{0.15em}
\noindent\textbf{Input representations.}
Re-indexing spectrograms onto a log-frequency axis is widely used in music signal processing~\cite{muller2021fmp}, while the CQT~\cite{brown1991cqt} additionally provides approximately constant relative frequency resolution.
HCQT-based representations have been widely adopted for MPE since Deep Salience~\cite{bittner2017deep}, including the Late/Deep CNN~\cite{cuesta2020multif0} and subsequent voice-assignment models~\cite{cuesta2022voiceassign}.
Basic Pitch~\cite{bittner2022lightweight} reduces the computational cost by constructing harmonic channels from frequency-shifted versions of a single CQT.
These representations are typically paired with convolutional architectures that preserve the pitch axis and exploit harmonic alignment.
Monophonic pitch estimators also commonly discretize pitch on a fine log-frequency grid~\cite{kim2018crepe}.

\begin{table*}[t]
\centering
\caption{
\textbf{A cappella ensembles and choirs.}
Each evaluation song is scored independently with multipitch accuracy, $TP/(TP+FP+FN)$, on the 360-bin, 20-cent pitch grid: exact-bin matching (20 cent) and a 50-cent tolerance.
The activation threshold is selected separately for each model to maximize the corresponding metric. Intervals denote the half-width of the 95\% bootstrap confidence interval over songs.
$^\ast$ChoralSynth is excluded from the training data of both our models and the published baselines.
}
\vspace{1ex}
\label{tab:heldout}
\small
\setlength{\tabcolsep}{1.2pt}
\begin{tabular}{@{}lcccccccccc@{}}
\toprule
& \multicolumn{2}{c}{jaCappella\,\cite{yoshii2023jacappella}} & \multicolumn{2}{c}{ACappellaSet\,\cite{pan2025acappellaset}} & \multicolumn{2}{c}{Korean\,\cite{aihub2021multisinger}} & \multicolumn{2}{c}{Cantor\'{\i}a\,\cite{cuesta2022cantoria}} & \multicolumn{2}{c}{ChoralSynth$^\ast$\,\cite{narang2023choralsynth}} \\
\cmidrule(lr){2-3}\cmidrule(lr){4-5}\cmidrule(lr){6-7}\cmidrule(lr){8-9}\cmidrule(lr){10-11}
Model & 20\,cent & 50\,cent & 20\,cent & 50\,cent & 20\,cent & 50\,cent & 20\,cent & 50\,cent & 20\,cent & 50\,cent \\
\midrule
HCQT~\cite{bittner2017deep} & & & & & & & & & & \\
\quad Late/Deep CNN~\cite{cuesta2020multif0} & .387$\pm$.039 & .804$\pm$.033 & .334$\pm$.026 & .716$\pm$.042 & .309$\pm$.008 & .791$\pm$.013 & .545$\pm$.012 & .863$\pm$.010 & .626$\pm$.020 & .932$\pm$.014 \\
\quad VoasCNN~\cite{cuesta2022voiceassign} & .291$\pm$.015 & .747$\pm$.040 & .208$\pm$.030 & .513$\pm$.074 & .255$\pm$.009 & .715$\pm$.020 & .358$\pm$.010 & .791$\pm$.022 & .469$\pm$.027 & .904$\pm$.018 \\
\quad VoasCLSTM~\cite{cuesta2022voiceassign} & .299$\pm$.018 & .731$\pm$.040 & .248$\pm$.018 & .573$\pm$.055 & .269$\pm$.008 & .745$\pm$.017 & .416$\pm$.011 & .819$\pm$.013 & .527$\pm$.029 & .894$\pm$.022 \\
\quad ConvNeXt~\cite{liu2022convnext} & .648$\pm$.036 & .884$\pm$.019 & .492$\pm$.027 & .828$\pm$.023 & .584$\pm$.016 & .880$\pm$.009 & .561$\pm$.017 & .875$\pm$.012 & .638$\pm$.020 & .934$\pm$.016 \\
\quad Harmonic stacking~\cite{bittner2022lightweight} & .649$\pm$.036 & .884$\pm$.019 & .496$\pm$.028 & .828$\pm$.024 & .585$\pm$.016 & .880$\pm$.009 & .562$\pm$.017 & .876$\pm$.011 & .640$\pm$.020 & .935$\pm$.016 \\
\midrule
Linear STFT & & & & & & & & & & \\
\quad ConvNeXt~\cite{liu2022convnext} & .683$\pm$.031 & .909$\pm$.013 & .533$\pm$.031 & .860$\pm$.019 & .614$\pm$.017 & .901$\pm$.008 & .563$\pm$.021 & .891$\pm$.011 & .657$\pm$.019 & .947$\pm$.014 \\
\quad TCN~\cite{bai2018tcn} & .694$\pm$.030 & .914$\pm$.013 & \textbf{.548$\pm$.033} & \textbf{.867$\pm$.017} & \textbf{.627$\pm$.016} & \textbf{.907$\pm$.007} & .597$\pm$.011 & .910$\pm$.005 & .642$\pm$.021 & .945$\pm$.013 \\
\quad Conformer~\cite{gulati2020conformer} & \textbf{.706$\pm$.028} & \textbf{.919$\pm$.011} & .544$\pm$.033 & .862$\pm$.019 & .621$\pm$.017 & .899$\pm$.008 & \textbf{.598$\pm$.010} & \textbf{.911$\pm$.006} & .631$\pm$.020 & .945$\pm$.012 \\
\quad ConvNeXt V2~\cite{woo2023convnextv2} & .681$\pm$.030 & .910$\pm$.013 & .537$\pm$.035 & .864$\pm$.017 & .619$\pm$.018 & .903$\pm$.007 & .587$\pm$.015 & .906$\pm$.008 & .648$\pm$.019 & .948$\pm$.013 \\
\quad InceptionNeXt~\cite{yu2024inceptionnext} & .692$\pm$.029 & .912$\pm$.012 & .536$\pm$.029 & .864$\pm$.018 & .618$\pm$.017 & .903$\pm$.008 & .564$\pm$.020 & .891$\pm$.011 & \textbf{.659$\pm$.019} & \textbf{.949$\pm$.013} \\
\bottomrule
\end{tabular}
\end{table*}

\section{Method}
\label{sec:method}

\subsection{Preliminaries: Frequency Resolution}
\label{sec:resolution}
Vocal-ensemble MPE requires distinguishing closely spaced fundamental frequencies.
Because musical intervals correspond to frequency ratios, their absolute spacing in Hz increases with frequency.
The Short-Time Fourier Transform(STFT) provides fixed frequency resolution determined by a single analysis window, whereas the CQT uses frequency-dependent analysis windows to maintain approximately constant relative frequency resolution across frequencies.
HCQT further stacks CQTs evaluated at harmonic ratios to align harmonic structure across pitch.
We investigate whether this frequency-adaptive analysis is necessary for accurate vocal-ensemble MPE by comparing it with fixed-resolution STFT representations.

\subsection{Pitch Quantization}
\label{sec:quant}
We formulate Multi Pitch Estimation (MPE) as frame-wise multi-label classification over a 20-cent pitch grid.
Following the 20-cent pitch discretization used in CREPE~\cite{kim2018crepe}, we use $K=360$ output bins spanning six octaves.
The center frequency of bin $k$ is
\begin{equation}
f_k = f_0 \cdot 2^{20k/1200}\ \text{Hz}, \quad
f_0 = 32.70\,\text{Hz}, \quad
k = 0,\dots,K-1 .
\end{equation}

The input representation is a fixed-resolution STFT with a window size of 2048 samples, a hop size of 256, and a sampling rate of 44.1\,kHz.
We provide the first $K$ STFT bins directly to the model,
\begin{equation}
\mathbf{G}[t,k] =
\log\!\left(1+\left|\mathbf{X}[t,k]\right|\right),
\quad k=0,\dots,K-1 .
\end{equation}
The logarithm compresses the large dynamic range between strong low-order partials and weak upper harmonics, and the added one keeps silent bins finite.
The resulting linear STFT spans 0 to 7.7\,kHz in uniform 21.5\,Hz intervals.
Unlike HCQT, its input bins are not pitch-aligned and its frequency resolution is fixed, and its spectral range extends beyond the predicted pitch range to include upper-harmonic content, as illustrated in Fig.~\ref{fig:concept}.

\subsection{Model}
\label{sec:arch}
As shown in Fig.~\ref{fig:arch}, two $3{\times}3$ convolutions produce 64-channel features, followed by a $1{\times}1$ projection, and a sigmoid output layer predicts activation over the 360 pitch bins.
The encoder operates along the temporal dimension and is varied across architectures.
We evaluate ConvNeXt~\cite{liu2022convnext}, TCN~\cite{bai2018tcn}, Conformer~\cite{gulati2020conformer}, ConvNeXt V2~\cite{woo2023convnextv2}, and InceptionNeXt~\cite{yu2024inceptionnext}.
All encoders use a hidden dimension of 256 and contain 18.5 to 19.4 million trainable parameters, with architecture-specific depths.
Following Basic Pitch~\cite{bittner2022lightweight}, we use binary targets with one positive bin per active pitch and optimize binary cross-entropy.
Predictions are decoded by thresholding the activation rather than by peak-picking a salience map. We use AdamW~\cite{loshchilov2019adamw} with a learning rate of $10^{-3}$, weight decay of $10^{-4}$, five warmup epochs, gradient clipping at 1.0, mixed precision, and cosine learning-rate decay over 100 epochs.

\begin{table}[t]
\centering
\caption{
\textbf{Training datasets.}
Training and validation splits are defined separately for each dataset. Hours denote the duration of songs.
}
\vspace{1ex}
\label{tab:data}
\small
\setlength{\tabcolsep}{3pt}
\begin{tabular}{@{}lccccl@{}}
\toprule
& & \multicolumn{2}{c}{Training} & \multicolumn{2}{c}{Validation} \\
\cmidrule(lr){3-4}\cmidrule(lr){5-6}
Dataset & Language & Songs & Hours & Songs & Hours \\
\midrule
Korean~\cite{aihub2021multisinger} & Korean & 51 & 50.9 & 48 & 14.5\ \\
jaCappella~\cite{yoshii2023jacappella} & Japanese & 32 & 40.3 & 8 & 4.8 \\
ACappellaSet~\cite{pan2025acappellaset} & Chinese & 66 & 32.8 & 17 & 6.4 \\
Cantor\'{\i}a~\cite{cuesta2022cantoria} & Latin & 11 & 8.2 & 3 & 0.9 \\
Dagstuhl~\cite{rosenzweig2020dagstuhlchoirset} & German & 6 & 0.8 & 2 & 0.2 \\
ESMUC~\cite{cuesta2022esmuc} & Various & 7 & 0.2 & 2 & 0.2  \\
CSD~\cite{cuesta2018csd} & Catalan & 2 & 0.1 & 1 & 0.1  \\
\midrule
\textbf{Total} & &\textbf{175} & \textbf{133.3} & \textbf{81} & \textbf{27.0} \\
\bottomrule
\end{tabular}
\end{table}

\begin{table}[t]
\centering
\caption{
\textbf{Choirs recorded in one room.}
Dagstuhl ChoirSet and ESMUC are the only evaluation datasets whose singers were recorded together, and both are included in the training data of the Late/Deep CNN and our models.
}
\vspace{1ex}
\label{tab:choirs}
\small
\setlength{\tabcolsep}{1.6pt}
\begin{tabular}{@{}lcccc@{}}
\toprule
& \multicolumn{2}{c}{Dagstuhl~\cite{rosenzweig2020dagstuhlchoirset}} & \multicolumn{2}{c}{ESMUC~\cite{cuesta2022esmuc}} \\
\cmidrule(lr){2-3}\cmidrule(lr){4-5}
& 20\,cent & 50\,cent & 20\,cent & 50\,cent \\
\midrule
HCQT & & & & \\
\quad Late/Deep CNN & .470$\pm$.012 & .904$\pm$.004 & \textbf{.564$\pm$.028} & \textbf{.910$\pm$.005} \\
\quad VoasCNN & .374$\pm$.015 & .871$\pm$.003 & .381$\pm$.018 & .862$\pm$.007 \\
\quad VoasCLSTM & .394$\pm$.013 & .858$\pm$.003 & .406$\pm$.024 & .841$\pm$.010 \\
\quad ConvNeXt & .617$\pm$.009 & .907$\pm$.010 & .475$\pm$.038 & .826$\pm$.011 \\
\quad Harmonic stacking & .602$\pm$.012 & .903$\pm$.007 & .485$\pm$.027 & .828$\pm$.017 \\
\midrule
Linear STFT & & & & \\
\quad ConvNeXt & .640$\pm$.010 & \textbf{.923$\pm$.003} & .480$\pm$.036 & .820$\pm$.018 \\
\quad TCN & \textbf{.643$\pm$.014} & .922$\pm$.004 & .498$\pm$.040 & .842$\pm$.017 \\
\quad Conformer & .576$\pm$.020 & .904$\pm$.005 & .491$\pm$.035 & .821$\pm$.024 \\
\quad ConvNeXt V2 & .641$\pm$.009 & \textbf{.923$\pm$.006} & .499$\pm$.037 & .840$\pm$.021 \\
\quad InceptionNeXt & .633$\pm$.007 & .919$\pm$.006 & .474$\pm$.044 & .829$\pm$.009 \\
\bottomrule
\end{tabular}
\end{table}

\section{Experiments}
\label{sec:setup}

\subsection{Experimental Setup}

\vspace{0.15em}
\noindent\textbf{Training data.}
The training data consist of isolated vocal stems from the datasets summarized in Table~\ref{tab:data}.
Reference $F_0$ is extracted from each stem before mixing with pYIN~\cite{mauch2014pyin} (10-cent resolution), except for Cantor\'{\i}a (RMVPE~\cite{wei2023rmvpe}) and Dagstuhl, ESMUC and CSD (the distributed references).
Training examples contain one to five within-song or cross-song stems, with 976 monophonic examples accounting for 22.5\% of the training set.
Each stem is independently scaled by a random gain within $\pm6$\,dB.
Two augmentations are then applied to each segment, each independently with probability 0.5: convolution with a synthetic room impulse response, with RT60 sampled from 0.1 to 0.8\,s and a dry ratio from 0.3 to 0.7, and resampling by up to $\pm2$ semitones, with the reference $F_0$ adjusted accordingly.

\vspace{0.15em}
\noindent\textbf{Evaluation sets.}
We evaluate on held-out songs from jaCappella~\cite{yoshii2023jacappella}, ACappellaSet~\cite{pan2025acappellaset}, Korean Multi-Singer, Cantor\'{\i}a~\cite{cuesta2022cantoria}, Dagstuhl~\cite{rosenzweig2020dagstuhlchoirset}, and ESMUC~\cite{cuesta2022esmuc}. CSD~\cite{cuesta2018csd} has a single held-out song and is not reported separately.
ChoralSynth~\cite{narang2023choralsynth} is the only evaluation dataset not represented in training and is also excluded from the training data of the compared models.

\vspace{0.15em}
\noindent\textbf{Metrics.}
The model outputs a sigmoid activation for each of the 360 pitch bins at every frame, which is converted to a binary prediction using an activation threshold.

\vspace{0.15em}
\noindent\textit{Song-level comparison metric.}
Following the Late/Deep CNN~\cite{cuesta2020multif0}, inference and evaluation are performed on complete songs and the activation threshold is selected separately for each model.
We compute multipitch accuracy, $TP/(TP+FP+FN)$, directly on the shared 20-cent pitch grid, under exact-bin matching and under a 50-cent tolerance, implemented on the grid as a two-bin offset that approximates the default in \texttt{mir\_eval}~\cite{raffel2014mireval}.

\vspace{0.15em}
\noindent\textit{Segment-level ablation metric.}
For the ablations we fix the threshold at 0.5 and compute exact-bin F1 on the 20-cent pitch grid for each 30\,s validation segment, averaged over 2{,}894 segments, which are also the resampling unit of the paired bootstrap intervals reported below.
Because exact-bin F1 is limited by the output-grid resolution, the window analysis is additionally scored against the reference $F_0$ before 20-cent quantization.
Local activation maxima are refined using the activation-weighted mean of each bin and its two neighboring bins, and are matched one-to-one with the references in each frame.
We report F1 at 10-, 20-, and 50-cent tolerances and the RMSE and MAE of the pitch error in cents for predictions matched within 50 cents.

\begin{figure}[t]
\centering
\includegraphics[width=\columnwidth]{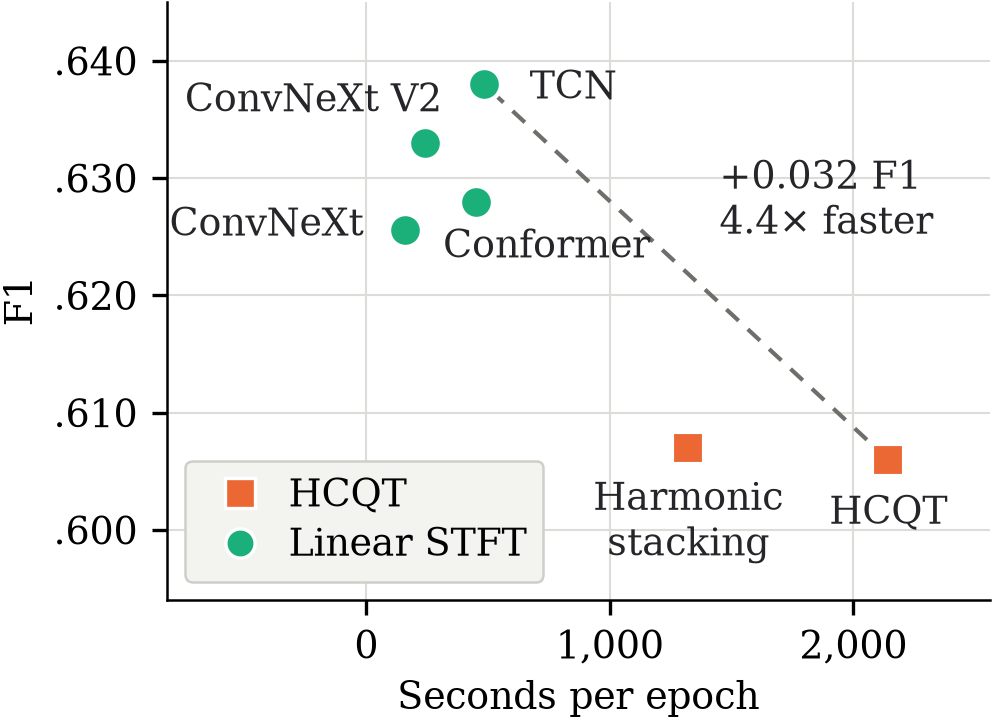}
\caption{
\textbf{F1 and computational cost.} Time per epoch re-timed on one A100. All models are trained and evaluated on 30\,s segments. InceptionNeXt coincides with ConvNeXt.
}
\label{fig:throughput}
\end{figure}

\begin{figure}[t]
\centering
\includegraphics[width=\columnwidth]{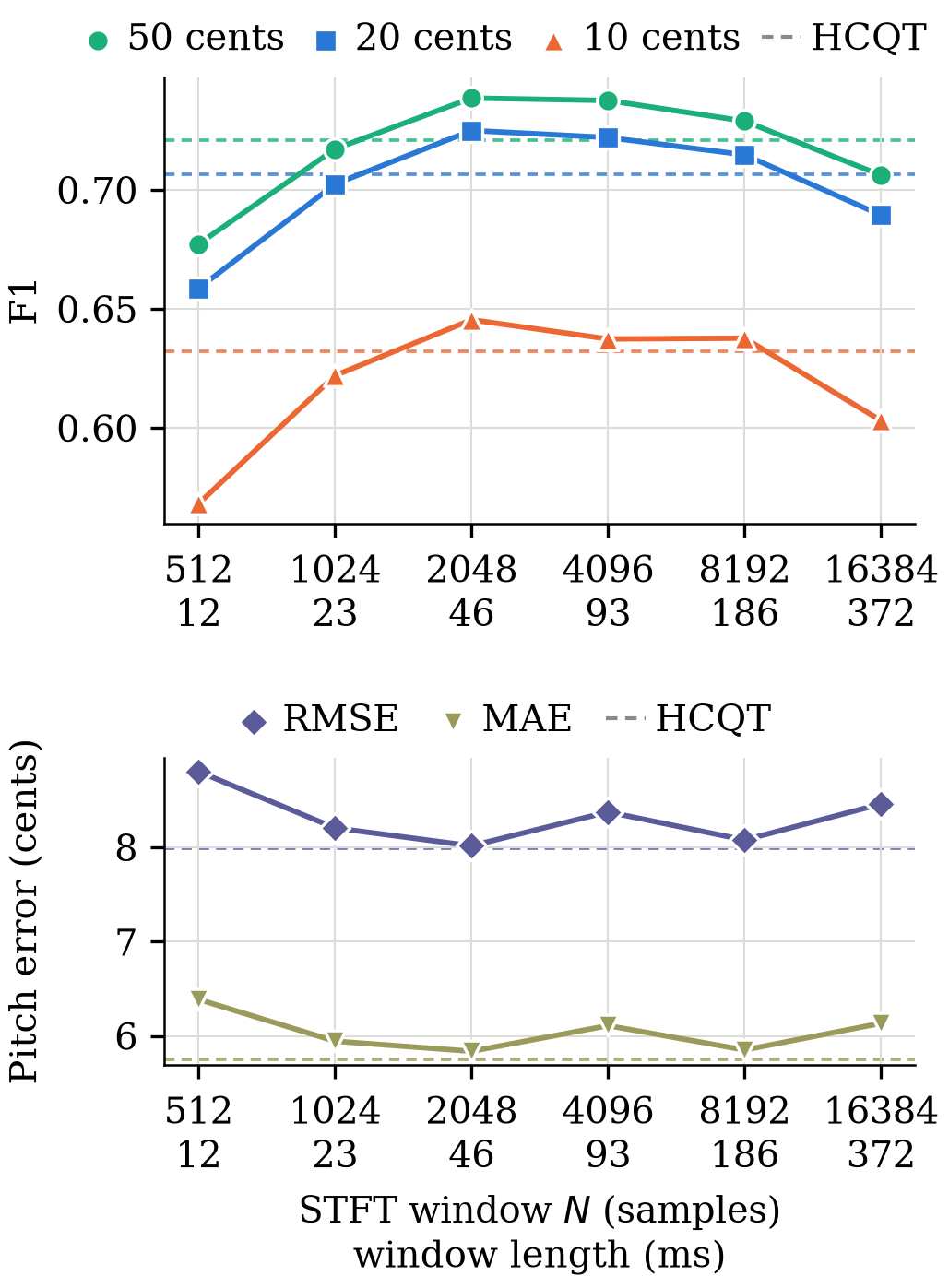}
\caption{
\textbf{Analysis window.}
Linear STFT with the ConvNeXt encoder across analysis-window sizes, scored against the reference $F_0$ before 20-cent quantization. Dashed lines: HCQT with the same encoder.
The top panel reports F1 at three pitch tolerances, and the bottom panel reports the pitch error of matched pitches.
}
\label{fig:window}
\end{figure}

\subsection{Results}
\label{sec:main}

\vspace{0.15em}
\noindent\textbf{A Cappella Ensembles and Choirs.}
Across encoder architectures, the linear STFT achieves higher F1 than the HCQT-based inputs while requiring substantially less computation, as shown in Fig.~\ref{fig:throughput}.
For the encoder comparison, the absolute F1 difference is 0.020 when pooled over all validation segments, with a 95\% paired bootstrap interval of 0.018 to 0.022.
This advantage extends to the song-level evaluation in Table~\ref{tab:heldout}, where the linear STFT achieves higher exact-bin accuracy on all five datasets, although the difference is smallest on Cantor\'{\i}a and generally decreases under the 50-cent tolerance.
The linear STFT also achieves higher accuracy than the Late/Deep CNN~\cite{cuesta2020multif0} and the two Voas models~\cite{cuesta2022voiceassign}, whose four voice-specific outputs are merged by maximum activation, on the same five datasets.
However, this pattern is less consistent on Dagstuhl and ESMUC, the two evaluation datasets in which the singers were recorded jointly, as reported in Table~\ref{tab:choirs}.

\vspace{0.15em}
\noindent\textbf{Frequency resolution and spectral coverage.}
Fig.~\ref{fig:window} varies the STFT analysis window with the spectral range fixed at 0--7.7\,kHz and the hop fixed at 256 samples.
F1 is highest with the 46\,ms window at all three tolerances and decreases for both shorter and longer windows, so trading frequency resolution for temporal resolution, or the reverse, does not improve performance.
Extending the spectral range beyond 7.7\,kHz yields no further improvement, whereas restricting it to C7, the upper bound of the predicted pitch range, reduces the advantage of the linear STFT, which associates the advantage with upper-harmonic content within 0--7.7\,kHz.
HCQT reaches a pitch error comparable to the best STFT configuration but a lower F1 at every tolerance, indicating that localization precision alone does not explain the performance difference between the representations.

\subsection{Limitations}
\label{sec:limits}
Reference $F_0$ is obtained automatically, mostly with pYIN, and its errors may propagate to the evaluation.
Most training and in-domain mixtures combine independently recorded stems of real singing and do not reflect the acoustic and tuning interactions present in jointly recorded ensembles.

\section{Conclusion}
\label{sec:conclusion}
We compared HCQT with a linear STFT that uses fixed frequency resolution and no pitch-aligned input grid.
Under a common encoder, the linear STFT outperforms HCQT while substantially reducing computational cost, and every additional encoder also exceeds HCQT in pooled F1.
Longer analysis windows and a spectral range beyond 7.7\,kHz do not improve performance, whereas limiting the input to the predicted pitch range reduces the advantage of the linear STFT.
These results indicate that a linear STFT provides an effective and computationally efficient input representation for vocal-ensemble MPE.
Future work will investigate whether these findings extend to musical instruments with different spectral and temporal characteristics.

\section{Acknowledgments}
This research was supported by the Basic Science Research Program through the National Research Foundation of Korea (NRF) funded by the Ministry of Education (RS-2025-25422688) and an IITP grant funded by the Korean Government (MSIT) (No. RS-2020-II201361, Artificial Intelligence Graduate School Program, Yonsei University).

{\normalsize
\let\oldthebibliography\thebibliography
\renewcommand{\thebibliography}[1]{\oldthebibliography{#1}%
  \setlength{\itemsep}{0pt}\setlength{\parskip}{0pt}}
\bibliographystyle{IEEEbib}
\bibliography{references}
}

\end{document}